\documentclass[12pt,preprint]{aastex}

\newcommand{\ks}{\mbox{\boldmath$\xi$}}
\newcommand{\nab}{{\bf \nabla}}
\newcommand{\GG}{\mbox{\bf g}}
\newcommand{\II}{\mbox{\bf I}}
\newcommand{\qq}{\mbox{\bf Q}}
\newcommand{\ee}{\hat{{\bf e}}_r}
\newcommand{\ez}{\hat{{\bf e}}_z}
\newcommand{\BB}{\mbox{\bf B}}
\newcommand{\vv}{\mbox{\bf v}}
\newcommand{\kk}{\mbox{\bf k}}
\newcommand{\ch}{\hat{\chi}}
\newcommand{\et}{\hat{\eta}}
\newcommand{\ze}{\hat{\zeta}}
\newcommand{\gm}{G_c M_*}
\newcommand{\vg}{V_g}
\newcommand{\ot}{\tilde{\omega}}

\slugcomment{Submitted to ApJL February 8th; Accepted March 13th} 

\shorttitle{MHD Spectroscopy of Accretion Disks}
\shortauthors{Keppens et al.}

\begin{document}


\title{Waves and Instabilities in Accretion Disks: \\ MHD Spectroscopic Analysis}


\author{R. Keppens, F. Casse, and J.P. Goedbloed}
\affil{FOM Institute for Plasma Physics Rijnhuizen, \\
P.O. Box 1207, 3430 BE Nieuwegein, The Netherlands \\
keppens@rijnh.nl, fcasse@rijnh.nl, goedbloed@rijnh.nl}



\begin{abstract}
A complete analytical and numerical treatment of all
magnetohydrodynamic waves and instabilities for radially stratified,
magnetized accretion disks is presented. The instabilities are a
possible source of anomalous transport. 
While recovering results on known hydrodynamic
and both weak- and strong-field magnetohydrodynamic perturbations, the
full magnetohydrodynamic spectra for 
a realistic accretion disk model demonstrates a much richer variety
of instabilities accessible to the plasma than previously realized.
We show that both weakly
and strongly magnetized accretion disks are prone to strong
non-axisymmetric instabilities.
The ability to characterize all waves arising in
accretion disks holds great promise for magnetohydrodynamic spectroscopic
analysis.
\end{abstract}




\section{Hydromagnetic stability for stationary equilibria}

Magnetohydrodynamic (MHD) spectroscopy~\citep{spect} 
entails the ability to calculate all MHD perturbations 
accessible to a particular magnetized plasma configuration (the 
{\em forward}
spectral problem), and holds the promise to use the MHD spectrum to diagnose 
the internal plasma state (the {\em backward} spectral problem).
In this letter, we predict all MHD waves and
instabilities for magnetized accretion disks. 

To analyse the entire MHD spectrum of gravitationally
and thermally
stratified, rotating, magnetized equilibrium configurations, we use
the formalism of~\citet{frrot}.
The equation of motion for the Lagrangian displacement $\ks$
of a fluid element is 
\begin{equation}
\rho\frac{\partial^2 \ks}{\partial t^2} + 2 \rho \vv\cdot\nab \frac{\partial \ks}{\partial t} + \nab \Pi - \BB\cdot \nab \qq - \qq \cdot \nab \BB
+ \nab \cdot (\rho \ks) \GG - \nab \cdot \left[ \rho \ks (\vv\cdot\nab)\vv -
\rho \vv \vv \cdot \nab \ks \right] = 0,
\label{fr}
\end{equation}
where $\Pi = -\ks \cdot \nab p -\gamma p \nab\cdot\ks + \BB\cdot \qq$ measures 
the Eulerian perturbation of total pressure,
$\qq = \nab \times (\ks \times \BB)$ is the Eulerian perturbation of the
magnetic field $\BB$, and $\GG$ is the gravitational
acceleration. Equation~(\ref{fr}) describes all waves
supported by a time-invariant equilibrium with
density $\rho$,
pressure $p$, flow field $\vv$, gravity $\GG$,
and $\BB$; $\gamma$ is the ratio of specific heats.
We specify our analysis to one-dimensional (1D)
axisymmetric stationary equilibria satisfying
\begin{equation}
\left(p + \frac{B^2_\theta+B_z^2}{2}\right)'=\rho\left(\frac{v_\theta^2}{r}-\frac{\gm}{r^2}\right)-\frac{B^2_\theta}{r},
\label{eq}
\end{equation}
where the prime denotes differentiation with
respect to the radial coordinate 
$r$ from a $(r,\theta,z)$ cylindrical system.
Hence, we concentrate on wave motions about an
equilibrium where centrifugal forces, gravity, pressure gradients and magnetic
forces balance. 
The one-dimensionality limits the generality by representing the gravitational
influence of a central body of mass $M_*$
by a line source $\GG (r)=-{\gm}/{r^2}\ee$, 
but this at most introduces a vertical wavelength cut-off on the validity
of our results. 

The equilibrium symmetry allows for a decoupled analysis of individual normal
mode solutions where $\ks(r,\theta,z,t)=
\left(\xi_r(r),\xi_\theta(r),\xi_z(r)\right)\exp i(m\theta+k z-\omega t)$. For the radially varying
magnetized equilibria, it is instructive to use a field line projection which
introduces the basic variables $\ch = r \xi_r$, $\et = i r \left(\BB/B \times \ee\right)\cdot \ks$ and $\ze = i r \BB/B \cdot \ks$.
It is then a matter of algebra to turn the Frieman-Rotenberg
system~(\ref{fr}) into 
\begin{eqnarray}
\left[
\left( \begin{array}{lcr}
r D_r \frac{\gamma p + B^2}{r} D_r -F^2 -r\left(\frac{B_\theta^2}{r^2}\right)' 
& 
r D_r \frac{\gamma p + B^2}{r} \frac{G}{B} - 2 k \frac{B_\theta B}{r}
&
r D_r \frac{\gamma p}{r}\frac{F}{B} \\
-\frac{G}{B} \left(\gamma p + B^2\right) D_r - 2 k \frac{B_\theta B}{r} 
&
- \left(\gamma p + B^2\right) \frac{G^2}{B^2} - F^2
&
- \gamma p \frac{GF}{B^2} \\
-\gamma p \frac{F}{B} D_r 
&
-\gamma p \frac{GF}{B^2} 
&
- \gamma p \frac{F^2}{B^2}
\end{array} \right)
\right.
& & \nonumber \\
\left.
+
\left( \begin{array}{lcr}
3 \frac{\gm}{r^3}\rho 
&
- \rho \vg \frac{G}{B}
&
- \rho \vg \frac{F}{B} \\
- \rho \vg \frac{G}{B}
&
0
&
0 \\
- \rho \vg \frac{F}{B}
&
0
&
0 
\end{array} \right)
+ 2 \rho \frac{v_\theta}{r} \ot \left( \begin{array}{lcr}
0
&
-\frac{B_z}{B}
&
-\frac{B_\theta}{B} \\
-\frac{B_z}{B}
&
0
&
0 \\
-\frac{B_\theta}{B} 
&
0
&
0 
\end{array} \right)
+\rho \ot^2 \II \right]
\left( \begin{array}{c}
\ch \\
\et \\
\ze
\end{array} \right)
& = 0. & 
\label{s3}
\end{eqnarray}
In this system, the radial derivative operator is denoted by $D_r$, while
the parallel gradient operator becomes the algebraic factor
$F\equiv -i \BB \cdot \nab = m B_\theta/r + k B_z$. 
Furthermore, 
$G \equiv m B_z/r - k B_\theta$ and $\vg \equiv v_\theta^2/r - \gm/r^2$.
The latter measures the deviation from a Keplerian disk.
The terms linearly proportional to the
Doppler shifted frequency $\ot = \omega - m v_\theta/r - k v_z$ 
represent the Coriolis effect. For static cylindrical equilibria without
external gravitational field, this representation of the eigenvalue problem
was introduced in~\citet{hans}.

From Equation~(\ref{s3}), one can immediately obtain the 
Alfv\'en and slow 
continuous parts of the ideal MHD spectrum by looking
at extreme localization
of the perturbations on single flux surfaces ($D_r \rightarrow \infty$). 
The first component can then be integrated at once,
and, when inserted in the second and third, yields
$A \et =0$ and $S \ze = 0$, respectively. The Alfv\'en continuum corresponds
to singular solutions $\et \propto \delta( r- r_A)$, where $A(r_A)\equiv
(\rho \ot^2 - F^2)(r_A)=0$. The slow modes are 
given by $\ze \propto \delta( r- r_S)$, where $S(r_S)\equiv
\left(\rho \ot^2 (\gamma p + B^2) - \gamma p F^2\right)(r_S)=0$. 
Due to the radial variation of the equilibrium and the Doppler shift, four
continuous ranges of real eigenfrequencies are found.
With the cluster points $\ot = \pm \infty$
for the fast subspectrum, these Doppler shifted forward and backward
Alfv\'en and slow continua 
determine the three-fold structure of the MHD spectrum.
These continua are not influenced by gravity 
or rotation.


An interesting limit of this system has obtained
a lot of attention in the accretion disk literature. Considering a
weakly magnetized accretion disk, 
it is possible to show that the combination
of differential rotation with a weak magnetic field introduces a
linear `weak-field shearing' MHD instability~\citep{bh}. As 
recently discussed in the review by~\citet{bhr}, one can derive 
a sixth order dispersion relation governing local linear
disturbances with purely vertical wave numbers $\kk = k \ez$ in disks
where the restricted hydrodynamic equilibrium relation $V_g=0$ holds.
This dispersion relation for axisymmetric $m=0$ modes is formally
found from
Equation~(\ref{s3}) by setting $D_r =0$, neglecting all curvature terms where
$B_\theta$ appears explicitly, and by using $V_g=0$.
It can then be written as 
\begin{eqnarray}
\left( \ot^2 -\frac{F^2}{\rho} \right) \left(
\ot^4 - k^2 \ot^2 (c_s^2 + \frac{B^2}{\rho}) + 
\frac{F^2}{\rho} k^2 c_s^2\right) & & \nonumber \\
- \left[ \kappa^2 \ot^4 - \ot^2 \left( k^2 \kappa^2 (c_s^2 + \frac{B_\theta^2}{\rho}) -
3 \frac{\gm}{r^3} \frac{F^2}{\rho} \right) \right] + 3 \frac{\gm}{r^3} \frac{F^2}{\rho} k^2 c_s^2  & = & 0.
\label{disp}
\end{eqnarray}
Here, $\kappa^2 \equiv 2 v_\theta (r v_\theta)'/r^2 = \gm/r^3$ is the
epicyclic frequency and the
squared sound speed is $c_s^2=\gamma p/\rho$. This
equation suggests marginal stability for ${F^2}/{\rho} = 3 {\gm}/{r^3}$.

An equivalent formulation of the MHD eigenvalue problem
for equilibria satisfying~(\ref{eq}) reduces the 
Frieman-Rotenberg system to a $2\times 2$ system in terms of the variables
$(\ch,\Pi)$:
\begin{equation}
\frac{A \, S}{r} \left( \begin{array}{c}
\ch \\
\Pi
\end{array} \right)'
+ \left( \begin{array}{cc}
C & D \\
E & -C
\end{array} \right)
\left( \begin{array}{c}
\ch \\
\Pi
\end{array} \right) = 0.
\label{s2}
\end{equation}
In this equation, the following terms appear
\begin{eqnarray}
C & = & \frac{\rho}{r}\vg \rho \ot^2 A + \frac{2m}{r^3} \left(B_\theta F +
\rho v_\theta \ot \right) S - 2 \frac{\rho^2 \ot^3 B_\theta}{r^2}\left(B_\theta
\ot + v_\theta F\right), \\
D & = & \rho^2 \ot^4 - \left(\frac{m^2}{r^2} + k^2\right) S, \\
E & = & \frac{A \, S}{r}\left[ -\frac{A}{r} -\left(\frac{B_\theta^2}{r^2}\right)'
+ \frac{\rho}{r}\kappa^2 + \frac{\rho'}{r}\vg -\frac{\rho^2}{r}\vg^2\frac{A}{S}
+ 4 \rho^2 \ot \frac{B_\theta}{r^2}\vg \frac{B_\theta \ot + v_\theta F}{S}
\right] \nonumber \\
 & & 
-\frac{4}{r^4}\left[ \rho^2 \ot^2 B_\theta^2 (B_\theta \ot + v_\theta F)^2 - \left( \left[B_\theta^2 + \rho v_\theta^2\right] F + 2 \rho \ot v_\theta B_\theta\right) F S \right]. 
\end{eqnarray}
This system~(\ref{s2}) reduces to the normal mode picture by~\citet{app} for
static cylindrical equilibria without gravity. \citet{bond}
and~\citet{ham} derived 
this system for cylindrical equilibria with flow, without 
gravity. 
An analysis for gravitating, flowing, magnetized equilibria in planar
geometry is found in~\citet{bart}. 

The set of equations~(\ref{s2}) again demonstrates that the
continuous part of the MHD spectrum is found from $A=0$ and $S=0$. The ranges 
in frequency where $D=0$ are not part of the continuous spectrum: for details,
see~\citet{hans2}.
Either of the formalisms given by the set~(\ref{s3}), system~(\ref{s2}), or
the equivalent second order Sturm-Liouville type equation
can be used to analyse the MHD spectrum of 
magnetized accretion disks. In particular, assuming a radial variation
$\exp(ik_r r)$ we obtain a local dispersion relation given by
\begin{equation}
k_r^2 \frac{A^2 S^2}{r^2} + C^2 + D E = 0.
\label{dloc}
\end{equation}
Similar to the static case, or the more general one including 
flow~\citep{bond},
one can prove the proportionality $C^2+D E \propto A S$, so that 
the continua $A S=0$ can be factored out. This leaves a sixth order
polynomial in $\ot$ which governs all discrete local modes.
Again, as a formal limit obtained by nullifying terms where
$m$, $k_r$, or $B_\theta$ appears explicitly, and assuming a Keplerian
equilibrium $\vg=0$, this polynomial is the dispersion relation~(\ref{disp}). 
The more general relation~(\ref{dloc}) leads immediately to the following
marginal 
stability criterion governing axisymmetric perturbations when $B_\theta=0$
\begin{equation}
F^2\left\{
\left(k_r^2+k^2\right)\frac{F^2}{\rho}+k^2\left(\kappa^2 - \frac{4 v_\theta^2}{r^2}-\frac{\vg^2}{c_s^2}+\frac{\rho'}{\rho}\vg\right)\right\}=0.
\label{m0b0}
\end{equation}
The multiplicative factor $F^2=k^2 B^2_z$ highlights the essential magnetic
character of this criterion. For non-vanishing $B_\theta$, even the $m=0$
case of equation~(\ref{dloc}) allows for overstable or damped MHD waves.
The general relation~(\ref{dloc}) recovers various known results.
\citet{Terq} presented a
stability analysis of accretion disks with both radial and vertical
equilibrium variations, but a purely toroidal $B_\theta$
magnetic field. In particular, a stability criterion
for axisymmetric perturbations required
a parameter defined in their equation~(29) to be positive. 
If one considers only radial equilibrium variation, the same is found from
our dispersion relation~(\ref{dloc}).
\citet{Kim} analyzed the stability of a cold ($p=0$)
radially stratified disk. Our relation~(\ref{eq})
contains their equilibrium as a special case, and their results form a
subset of our analysis. E.g., their equation~(43) for
`poloidal buoyancy modes' follows
from~(\ref{dloc}) by taking $m=B_{\theta}=v_z=0$ and assuming
$k_z^2 B_{z}^2/\rho \gg \ot^2$.
As a final note on the system~(\ref{s2}), the pure hydrodynamical limit
can be written as
\begin{equation}
\left( \begin{array}{c}
\ch \\
\Pi
\end{array} \right)'
+ \left( \begin{array}{cc}
\frac{p'}{\gamma p} + 2 \frac{m v_\theta}{r^2 \ot} & r\frac{\ot^2 - \left(\frac{m^2}{r^2}+k^2\right) c_s^2}{\rho c_s^2 \ot^2} \\
\left(\frac{\rho'}{\rho}-\frac{p'}{\gamma p}\right)\frac{p'}{r} -\frac{\rho}{r}
\left(\ot^2 - \kappa^2\right) & -\frac{p'}{\gamma p} - 2 \frac{m v_\theta}{r^2 \ot}
\end{array} \right)
\left( \begin{array}{c}
\ch \\
\Pi
\end{array} \right) = 0.
\label{s2hd}
\end{equation}
The continuous part of the spectrum is now collapsed onto the flow
continuum $\ot^2=0$ \citep{case}. 
The local dispersion relation~(\ref{dloc}) 
with this flow continuum factored out reads
\begin{equation}
k_r^2c_s^2\ot^2 + c_s^2\left(2\frac{m v_\theta}{r^2}+\ot\frac{p'}{\gamma p}\right)^2
+\left[\ot^2 - \left(\frac{m^2}{r^2}+k^2\right)c_s^2\right]\left[\kappa^2 -\ot^2+\vg\left(\frac{\rho'}{\rho}-\frac{p'}{\gamma p}\right)\right] = 0.
\label{hddisp}
\end{equation}
Several known results are immediately recovered 
from this relation, especially for $m=0$ modes where it has a simple solution. 
For a polytropic equilibrium with $v_z=0$, the limit
where $(k_r, k) \rightarrow \infty$ at a finite ratio $k_r/k$ yields 
$\ot^2 = \kappa^2 k^2/(k_r^2 +k^2)$, giving a dense range of 
discrete modes within $-\kappa < \ot < \kappa$. The same limit shows that
stability requires $\kappa^2 > 0$, known as the~\citet{ray} criterion. 
The $m\neq 0$ global modes of equation~(\ref{hddisp}) are 
of the~\citet{pap} variety. 
High $m$ non-axisymmetric instabilities can be inferred from~(\ref{hddisp}), 
which simplifies
in the incompressible limit ($\gamma\rightarrow\infty$, $\rho'\rightarrow 0$) 
to a bi-squared relation.
For weakly magnetized disks, the modes contained
in the dispersion relation~(\ref{hddisp}) will be complemented by
Alfv\'enic and slow magnetosonic perturbations.

\section{Full spectral analysis of accretion disks}

With the general formalism for a complete normal mode analysis of
magnetized accretion disks in place, 
we present in the remainder 
full MHD spectra for a specific thin disk model satisfying~(\ref{eq}). 
We use 
an analytical model of a disk 
where the ratio $\beta=2p/B^2$ is $r$-independent. 
Two other parameters
are 
the constant helicity of the magnetic field $\alpha=
-B_{\theta}/B_z$,
and the disk aspect ratio $\varepsilon = H/r \ll 1$,
where $H$ is the disk scale height. If one neglects radiative
pressure in the disk, $\varepsilon$ measures the
ratio of the sound to the Keplerian speed
$\varepsilon=c_s/\Omega_k r$ \citep{Shak73}. 
Because of the Newtonian gravitational field, we
describe the physical quantities using radial power laws. 
To have a constant
$\varepsilon$, $c_s$ must scale as
$r^{-1/2}$. 
The density 
profile is set to $\rho\propto r^{-3/2}$.
For the pressure and magnetic field we use
the $r^{-1/2}$ scaling of $c_s$
and the constant $\beta$.
The inner radius,
the density and the Keplerian velocity at the inner
radius make all quantities dimensionless.
The resulting disk structure is: 
$\rho=r^{-3/2}$, 
$p=\varepsilon^2r^{-5/2}$, magnetic field $B_z=\varepsilon
\sqrt{2}r^{-5/4}/\sqrt{\beta(1+\alpha^2)}$ and $B_{\theta}=-\alpha B_z$. The rotational
velocity will then be $v_{\theta}=v_or^{-1/2}$, where 
$v_o(\alpha,\beta,\varepsilon)$ is found from~(\ref{eq}).
The $v_z$ component is set to zero.
We calculate
the collection of eigenfrequencies by a
generalization of the LEDAFLOW program~\citep{nijb}. 
LEDAFLOW computes the entire MHD spectrum of 1D,
gravitationally stratified, magnetized equilibria with flow by solving the
linearised MHD equations. 
The radial discretization for the eigenfunctions
employs quadratic and cubic Hermite finite elements, appropriate for
both global and (near-)singular local modes. We use
rigid boundary conditions in the domain
$r\in [1,r_{out}=10]$. 
We input the 
parameters for the equilibrium ($\alpha,\beta, \varepsilon$) and
the normal mode numbers ($m,k$). Since we consider thin disks, we set
$\varepsilon=0.1$.
The helicity of the $\BB$ field is taken $\alpha=1$. 

\begin{figure}[t]
\plotone{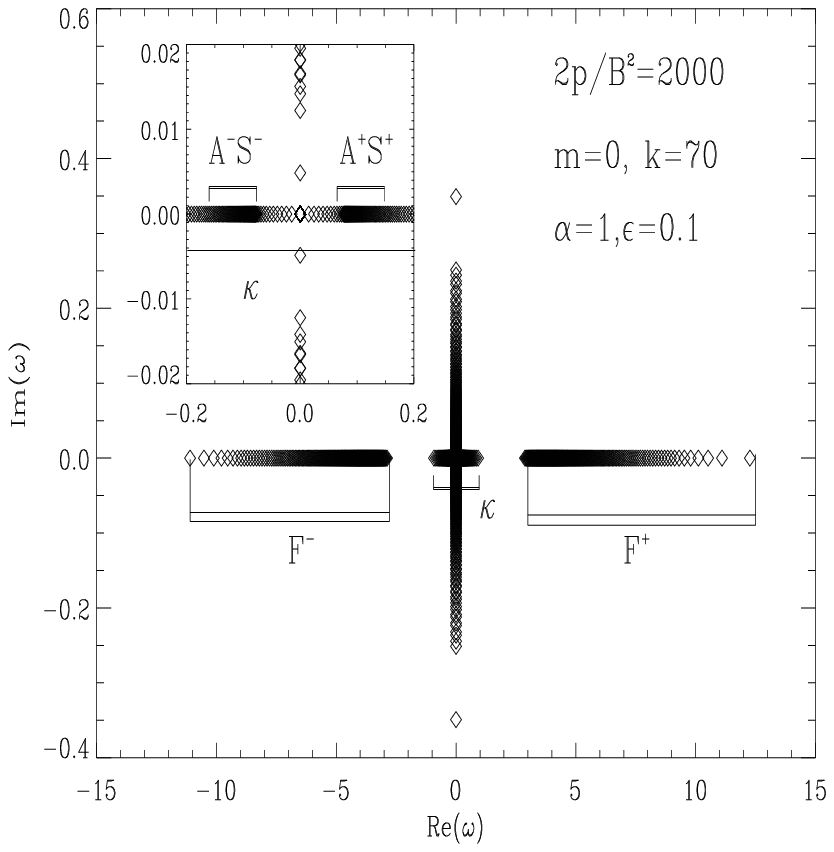}
\caption{MHD spectrum for a weakly magnetized accretion disk
for perturbations where ${\bf k}=k{\bf e}_z$. As the
Doppler shift vanishes (${\bf k.v}=0$), the identification of the different
subspectra is easy ($A^{\pm},S^{\pm}$ for forward and backward Alfv\'en and
Slow modes, $\kappa$ for the epicyclic ones, and $F^{\pm}$ for forward and
backward fast modes.).
The magneto-rotational instability is identified as a
cluster spectrum of discrete modes associated with the slow
continuum.\label{fig1}}
\end{figure}

\begin{figure}[t]
\plotone{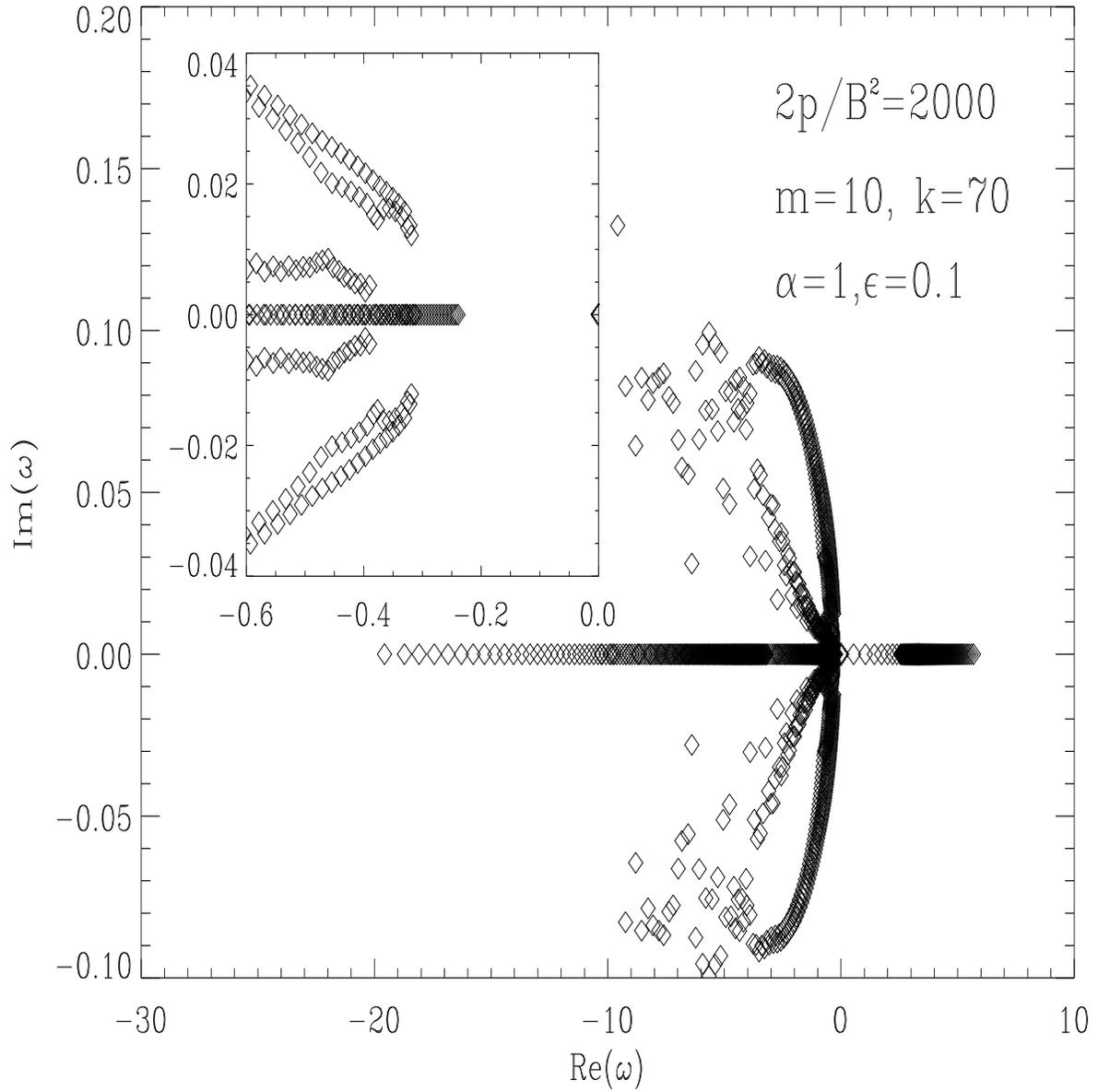}
\caption{As Fig.~(\ref{fig1}), but for non-axisymmetric
perturbations $(m=10)$. 
All the slow and Alfv\'enic sub-spectra
overlap partially and several cluster branches of unstable modes are apparent.\label{fig2}}
\end{figure}

\underline{{\em Weakly magnetized accretion disk.}} We present two
calculations  for a weakly magnetized accretion disk. We set the parameter
$\beta=2\ 10^3$. Similar to~\citet{bhr}, we first take a purely vertical
wave vector ${\bf k}=2\pi/H{\bf
e}_z$. The vertical wavelength should not exceed the
vertical size of the disk. In Fig.~(\ref{fig1}), we show
the MHD spectrum.
We can identify all modes in this plot: the fast-magnetosonic, epicyclic, and
(overlapping) Alfv\'en and slow-magnetosonic continuum modes are all
on the real axis. A dense sequence of discrete modes
appears on the imaginary axis: the magneto-rotational instability. 
This sequence is the
unstable part of the slow magnetosonic subspectrum.
Performing the same calculation for $\beta \rightarrow \infty$ 
(hydrodynamical case), only the epicyclic and
sonic modes remain and no unstable mode is found, confirming its
slow magnetosonic nature.
We present in
Fig.~(\ref{fig2}) the spectrum for non-axisymmetric perturbations
with $m=10,k=70$. Note the asymmetry due to the Doppler shift ${\bf k\cdot
v}\neq 0$. 
The identification of the different modes is quite involved since the
fast, slow, Alfv\'en and epicyclic frequency ranges are all overlapping. 
Various branches of unstable modes 
can be seen, where only the rightmost branches correspond to their
axisymmetric analogues already present in Figure~(\ref{fig1}).
The structure suggests
the presence of global modes accumulating towards both
the Alfv\'en and the slow continua. A full MHD spectroscopic analysis, which
investigates the mode types from their specific polarization properties,
is the subject of future work. 

\begin{figure}[t]
\plotone{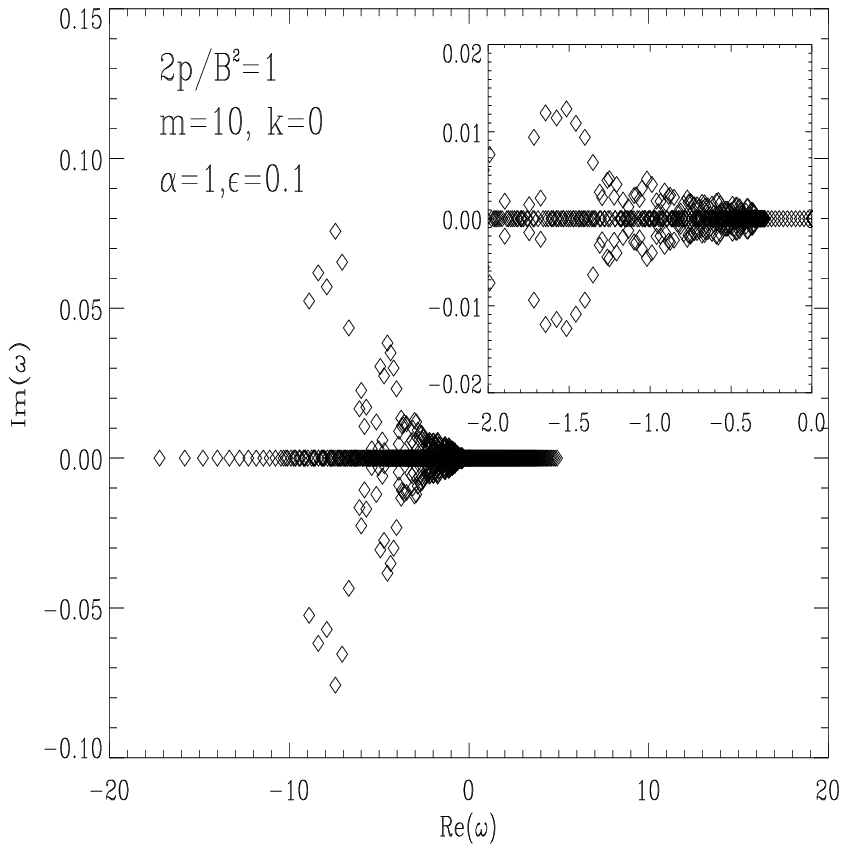}
\caption{MHD spectrum of toroidal perturbations in an equipartition
accretion disk ($\beta=1$). The calculation demonstrates a large collection
of unstable modes.\label{fig3}}
\end{figure}

\underline{{\em Equipartition accretion disk ($\beta=1$).}} We performed
several calculations of MHD perturbations 
supported by disks where the magnetic pressure is of the same order
as the thermal one. The results we have obtained are quite clear. If
the perturbation is non-axisymmetric, unstable branches appear in
the spectra. The growth rate of these
instabilities grows for increasing $m$. 
For the disk equilibrium discussed above, no axisymmetric
unstable modes are
found.
The spectral structure is 
shown in Fig.~(\ref{fig3}) for $m=10$.
Because all the
terms in Eq.~(\ref{s2}) are of the same order,
a purely analytical approach is virtually impossible.

\section{Concluding remarks}  

We applied
a general 
formalism to an astrophysical system where most MHD instabilities are believed
to occur: the accretion disk. Our analysis covers 
both unmagnetized and magnetized disks. Complete
MHD spectra were presented for equilibria involving
the gravity of a central object, velocities $v_{\theta}$ and $v_z$, 
and both thermal pressure and a magnetic field with azimuthal
and vertical components. 
We deduced the relations~(\ref{dloc}), (\ref{m0b0}),
(\ref{hddisp}), describing the threshold of several known instabilities for
a differentially rotating fluid, including the Papaloizou-Pringle modes and
the weak-field magneto-rotational one. The relations apply to fully
compressible, non-Keplerian magnetized disks. 
If one ignores vertical stratification, they generalize 
the analysis of~\citet{Papa}, who considered only $m=0$
modes when no azimuthal $B_\theta$ field was present.
We complemented our analytical
approach with numerical calculations of MHD spectra of a realistic magnetized
disk model. MHD spectra for weakly magnetized accretion
disks contain the magneto-rotational instability and several other
branches of discrete unstable modes. 
We numerically found a variety of
toroidal unstable modes which can affect equipartion accretion disks.
Our results are consistent
with~\citet{Nogu} who made an analysis of non-axisymmetric
incompressible modes in accretion disks, and found Alfv\'enic instabilities
for strong magnetic fields. We generalize their findings
by considering all MHD modes. Instabilities at equipartition field
strengths are of big interest since it is believed that magnetized
disks launching MHD jets have a thermal pressure of the same order
as the magnetic one~\citep{Ferr}. 
\citet{Coppi98,Coppi01}
have pointed out the importance of anomalous transport produced
by non-singular,
non-axisymmetric bending modes for the regime of large magnetic energy
density. These modes are amongst the instabilities analysed here.
A complete MHD spectroscopic study for these and other
(also 2D) accretion disk equilibria  will be
presented in future work~\citep{hanslet}. 
Since the MHD spectrum for disks which are
weakly magnetized or at equipartition field strengths shows this enormous
variety of unstable modes, the description of MHD
turbulence in magnetized accretion disks is far from complete.
Follow-up nonlinear simulations should reveal the
importance of the various modes for triggering and sustaining MHD turbulence. 
The analytical framework presented can be of interest for
physical mechanisms involving disk instabilities,
such as the `Accretion-Ejection
Instability'~\citep{Tagg1} and the hydrodynamical Rossby vortices~\citep{Tagg2}.
The study of instabilities
affecting radially stratified 
MHD jets~\citep{Kers,Appl,Kim} is embedded in the present 
formalism.

\acknowledgments

This work was done under Euratom-FOM Association Agreement with
financial support from NWO, Euratom, and the European Community's Human
Potential Programme under contract HPRN-CT-2000-00153, PLATON, also
acknowledged by F.C.

\end{document}